\documentclass[runningheads,a4paper]{llncs}

\makeatletter
\@ifundefined{lhs2tex.lhs2tex.sty.read}%
  {\@namedef{lhs2tex.lhs2tex.sty.read}{}%
   \newcommand\SkipToFmtEnd{}%
   \newcommand\EndFmtInput{}%
   \long\def\SkipToFmtEnd#1\EndFmtInput{}%
  }\SkipToFmtEnd

\newcommand\ReadOnlyOnce[1]{\@ifundefined{#1}{\@namedef{#1}{}}\SkipToFmtEnd}
\usepackage{amstext}
\usepackage{amssymb}
\usepackage{stmaryrd}
\DeclareFontFamily{OT1}{cmtex}{}
\DeclareFontShape{OT1}{cmtex}{m}{n}
  {<5><6><7><8>cmtex8
   <9>cmtex9
   <10><10.95><12><14.4><17.28><20.74><24.88>cmtex10}{}
\DeclareFontShape{OT1}{cmtex}{m}{it}
  {<-> ssub * cmtt/m/it}{}

\DeclareFontShape{OT1}{cmtt}{bx}{n}
  {<5><6><7><8>cmtt8
   <9>cmbtt9
   <10><10.95><12><14.4><17.28><20.74><24.88>cmbtt10}{}
\DeclareFontShape{OT1}{cmtex}{bx}{n}
  {<-> ssub * cmtt/bx/n}{}

\newcommand{\Conid}[1]{\mathit{#1}}
\newcommand{\Varid}[1]{\mathit{#1}}
\newcommand{\anonymous}{\kern0.06em \vbox{\hrule\@width.5em}}

\usepackage{polytable}

\@ifundefined{mathindent}%
  {\newdimen\mathindent\mathindent\leftmargini}%
  {}%

\def\resethooks{%
  \global\let\SaveRestoreHook\empty
  \global\let\ColumnHook\empty}
\newcommand*{\savecolumns}[1][default]%
  {\g@addto@macro\SaveRestoreHook{\savecolumns[#1]}}
\newcommand*{\restorecolumns}[1][default]%
  {\g@addto@macro\SaveRestoreHook{\restorecolumns[#1]}}
\newcommand*{\aligncolumn}[2]%
  {\g@addto@macro\ColumnHook{\column{#1}{#2}}}

\resethooks

\newcommand{\onelinecommentchars}{\quad-{}- }
\newcommand{\commentbeginchars}{\enskip\{-}
\newcommand{\commentendchars}{-\}\enskip}

\newcommand{\visiblecomments}{%
  \let\onelinecomment=\onelinecommentchars
  \let\commentbegin=\commentbeginchars
  \let\commentend=\commentendchars}

\newcommand{\invisiblecomments}{%
  \let\onelinecomment=\empty
  \let\commentbegin=\empty
  \let\commentend=\empty}

\visiblecomments

\newlength{\blanklineskip}
\newcommand{\hsindent}[1]{\quad}
\let\hspre\empty
\let\hspost\empty

\EndFmtInput
\makeatother
\ReadOnlyOnce{polycode.fmt}%
\makeatletter

\newcommand{\hsnewpar}[1]%
  {{\parskip=0pt\parindent=0pt\par\vskip #1\noindent}}

\newcommand{\hscodestyle}{}

\newcommand{\sethscode}[1]%
  {\expandafter\let\expandafter\hscode\csname #1\endcsname
   \expandafter\let\expandafter\endhscode\csname end#1\endcsname}

  {\par\noindent
   \advance\leftskip\mathindent
   \hscodestyle
   \let\\=\@normalcr
   \let\hspre\(\let\hspost\)%
   \pboxed}%
  {\endpboxed\)%
   \par\noindent
   \ignorespacesafterend}

  {\hsnewpar\abovedisplayskip
   \advance\leftskip\mathindent
   \hscodestyle
   \let\hspre\(\let\hspost\)%
   \pboxed}%
  {\endpboxed%
   \hsnewpar\belowdisplayskip
   \ignorespacesafterend}

  {\hsnewpar\abovedisplayskip
   \advance\leftskip\mathindent
   \hscodestyle
   \let\\=\@normalcr
   \(\pboxed}%
  {\endpboxed\)%
   \hsnewpar\belowdisplayskip
   \ignorespacesafterend}

\newcommand{\plainhs}{\sethscode{plainhscode}}

\plainhs

  {\hsnewpar\abovedisplayskip
   \advance\leftskip\mathindent
   \hscodestyle
   \let\\=\@normalcr
   \(\parray}%
  {\endparray\)%
   \hsnewpar\belowdisplayskip
   \ignorespacesafterend}

  {\parray}{\endparray}

  {\(\parray}{\endparray\)}

\def\codeframewidth{\arrayrulewidth}
\RequirePackage{calc}

  {\parskip=\abovedisplayskip\par\noindent
   \hscodestyle
   \arrayrulewidth=\codeframewidth
   \tabular{@{}|p{\linewidth-2\arraycolsep-2\arrayrulewidth-2pt}|@{}}%
   \hline\framedhslinecorrect\\{-1.5ex}%
   \let\endoflinesave=\\
   \let\\=\@normalcr
   \(\pboxed}%
  {\endpboxed\)%
   \framedhslinecorrect\endoflinesave{.5ex}\hline
   \endtabular
   \parskip=\belowdisplayskip\par\noindent
   \ignorespacesafterend}

\newcommand{\framedhslinecorrect}[2]%
  {#1[#2]}

  {\(\def\column##1##2{}%
   \let\>\undefined\let\<\undefined\let\\\undefined
   \newcommand\>[1][]{}\newcommand\<[1][]{}\newcommand\\[1][]{}%
   \def\fromto##1##2##3{##3}%
   }{\) }%

  {\let\orighscode=\hscode
   \let\origendhscode=\endhscode
   \def\endhscode{\def\hscode{\endgroup\def\@currenvir{hscode}\\}\begingroup}
   \orighscode\def\hscode{\endgroup\def\@currenvir{hscode}}}%
  {\origendhscode
   \global\let\hscode=\orighscode
   \global\let\endhscode=\origendhscode}%

\makeatother
\EndFmtInput
\ReadOnlyOnce{agda.fmt}%

\RequirePackage[T1]{fontenc}
\RequirePackage[utf8x]{inputenc}
\RequirePackage{ucs}
\RequirePackage{amsfonts}

\DeclareUnicodeCharacter{737}{\textsuperscript{l}}
\DeclareUnicodeCharacter{8718}{\ensuremath{\blacksquare}}
\DeclareUnicodeCharacter{8759}{::}
\DeclareUnicodeCharacter{9669}{\ensuremath{\triangleleft}}
\DeclareUnicodeCharacter{8799}{\ensuremath{\stackrel{\scriptscriptstyle ?}{=}}}
\DeclareUnicodeCharacter{10214}{\ensuremath{\llbracket}}
\DeclareUnicodeCharacter{10215}{\ensuremath{\rrbracket}}

\renewcommand\Varid[1]{\mathord{\textsf{#1}}}
\let\Conid\Varid
\newcommand\Keyword[1]{\textsf{\textbf{#1}}}
\EndFmtInput

\usepackage{ucs}
\usepackage{amssymb}
\usepackage[english]{babel}
\usepackage{graphicx}
\usepackage{hyperref}
\usepackage{authblk}
\usepackage{textgreek}
\usepackage[references]{agda}
\usepackage{url}
\newcommand{\keywords}[1]{\par\addvspace\baselineskip
\noindent\keywordname\enspace\ignorespaces#1}

\newcommand{\jkeyw}[1]{\texttt{\bfseries #1}}

\DeclareUnicodeCharacter{8336}{\_a}
\DeclareUnicodeCharacter{8337}{\_e}
\DeclareUnicodeCharacter{9657}{\triangleright}

\begin{document}

\mainmatter  

\title{Verified type checker for Jolie programming language}

\titlerunning{Verified type checker for Jolie programming language}

%
%
\author[1]{Evgenii Akentev}
\author[2]{Alexander Tchitchigin}
\author[1,3]{Larisa Safina}
\author[1]{Manuel Mazzara}

\affil[1]{Innopolis University}
\affil[$\relax$]{\email{ak3ntev@gmail.com}, \email{\{l.safina, m.mazzara\}@innopolis.ru}}
\affil[2]{Innosoft LLC}
\affil[$\relax$]{\email{sad.ronin@gmail.com}}
\affil[3]{University of Southern Denmark}
\authorrunning{Verified type checker for Jolie programming language}

\institute{Innopolis University}

%
%

\toctitle{Verified type checker for Jolie programming language}
\tocauthor{Authors' Instructions}
\maketitle

\begin{abstract}
Jolie is a service-oriented programming language which comes with the formal specication of its type system. However, there is no tool to ensure that programs in Jolie are well-typed. In this paper we provide the results of building a type checker for Jolie as a part of its syntax and semantics formal model. We express the type checker as a program with dependent types in Agda proof assistant which helps to ascertain that the type checker is correct.

\end{abstract}

\if{False}
\begin{hscode}\SaveRestoreHook
\column{B}{@{}>{\hspre}l<{\hspost}@{}}%
\column{E}{@{}>{\hspre}l<{\hspost}@{}}%
\>[B]{}\mbox{\onelinecomment  Imports}{}\<[E]%
\\
\>[B]{}\Keyword{open}\;\Keyword{import}\;\Conid{Data.String}\;\Keyword{using}\;(\Conid{String}){}\<[E]%
\\
\>[B]{}\Keyword{open}\;\Keyword{import}\;\Conid{Data.Integer}\;\Keyword{using}\;(\Conid{ℤ};\Varid{+\char95 }){}\<[E]%
\\
\>[B]{}\Keyword{open}\;\Keyword{import}\;\Conid{Data.Product}\;\Keyword{using}\;(\Varid{\char95 ×\char95 }){}\<[E]%
\\
\>[B]{}\Keyword{open}\;\Keyword{import}\;\Conid{Data.Maybe}\;\Keyword{using}\;(\Conid{Maybe}){}\<[E]%
\\
\>[B]{}\Keyword{open}\;\Keyword{import}\;\Conid{Data.List}\;\Keyword{using}\;(\Conid{List}){}\<[E]%
\\
\>[B]{}\Keyword{open}\;\Keyword{import}\;\Conid{Data.Bool}\;\Keyword{using}\;(\Conid{Bool}){}\<[E]%
\\
\>[B]{}\Keyword{open}\;\Keyword{import}\;\Conid{Data.Nat}\;\Keyword{using}\;(\Conid{ℕ}){}\<[E]%
\\
\>[B]{}\Keyword{open}\;\Keyword{import}\;\Conid{Data.List.All}\;\Keyword{using}\;(\Conid{All}){}\<[E]%
\\
\>[B]{}\Keyword{open}\;\Keyword{import}\;\Conid{Relation.Nullary}\;\Keyword{using}\;(\Varid{¬\char95 }){}\<[E]%
\\
\>[B]{}\Keyword{open}\;\Keyword{import}\;\Conid{Relation.Binary.PropositionalEquality}\;\Keyword{using}\;(\Varid{\char95 ≡\char95 };\Varid{refl}){}\<[E]%
\\
\>[B]{}\Keyword{open}\;\Keyword{import}\;\Conid{Data.Vec}\;\Keyword{using}\;(\Conid{Vec};\Varid{[]};\Varid{\char95 ++\char95 };\Varid{\char95 ∷\char95 };\Varid{here};\Varid{there}){}\<[E]%
\\
\>[B]{}\Keyword{open}\;\Keyword{import}\;\Conid{Data.Product}\;\Keyword{using}\;(\Varid{\char95 ,\char95 };\Varid{\char95 ×\char95 }){}\<[E]%
\\
\>[B]{}\Keyword{open}\;\Keyword{import}\;\Conid{Function}\;\Keyword{using}\;(\Varid{\char95 \$\char95 }){}\<[E]%
\ColumnHook
\end{hscode}\resethooks
\fi

\section{Introduction}

Microservices architecture is a modern paradigm in software development using a composition of autonomous entities for creating systems~\cite{DBLP:journals/corr/DragoniGLMMMS16}.
It has been developed as the answer to the problems arisen in applications built in monolith or Service-Oriented Architecture styles including difficulties with scalability, complexity and dependencies of the evolving application. Microservices implement only a limited and cohesive amount of functionality, run their own processes, and use lightweight communication mechanisms.

In the fast growing landscape of microservices, Jolie~\cite{montesi2010jolie} appears to be a good candidate to play the role of paradigmatic programming language~\cite{MGZ14}.
Since every program in Jolie is a microservice, everything can be reused or recomposed for obtaining new microservices making easy creation of as simple services as complex architectural compositions. This makes programs to scale easily, thus supports distributed architecture with simple managing of components, reducing maintenance and lower development costs.

However, communication between microservices in Jolie is obtained by means of sending and receiving messages whose types correspondence is checked only at runtime. Having a formalized type system of programming language gives us an opportunity to implement type checking mechanism and use it before the runtime.

Our idea is to augment Jolie with static type-checking mechanism based on type system specification. Type system of Jolie, described by Nielsen in ~\cite{nielsen2013type}, represents typing rules for the core of programming language (excluding subtyping, recursive types and some other primitives) and gives as a theoretical basis to reason about correctness of a program in Jolie.

We decided to use Agda ~\cite{bove,agda} as a proof assistant for implementing our type checker. Agda is a functional programming  based on dependent types. Agda represents ans extension to the Martin-L{\"o}f’s logical framework~\cite{lof,nordstrom90}. Thus we can introduce logical propositions as types by means of Curry-Howard isomorphism~\cite{Sorensen} and prove them writing the type corresponding programs. Agda possesses concrete syntax and comes with rich family of data types, pattern matching mechanism, termination checking, as well as with the ordinary programming constructs.

The paper has the following structure.
Section~\ref{sec:jolform} provides a background for Jolie programming language and  presents a the subset of its formalization\footnote{The whole formalization is available here~\cite{aken}} written in Agda. We implement the syntax of the behavioral layer of Jolie, which describes the workflow of service activities, in section~\ref{sec:jolform}, and in section~\ref{sec:typesys} we provide the necessary subset of typing rules.
Section~\ref{sec:strcongr} contains the proof of "Structural Congruence" lemma for behaviours showing the correctness of rules introduced.
Finally, in section~\ref{sec:concl} we conclude our paper and describe the possible directions of future work.

\section{Jolie Formalization}\label{sec:jolform}
Formal syntax and semantic of Jolie are based on SOCK process calculi~\cite{Gui07,Guidi}.
SOCK was created for designing service-oriented systems and was inspired by notable $\pi$-calculus \cite{Milner1999} and WS-BPEL~\cite{BPEL}.
Primitives in SOCK are able to express one-way and request-response communication,  parallel and sequential behavior of processes and control primitives.

SOCK (so a formalization of Jolie program) comprises of three layers:

\begin{itemize}
\item \textit{Behavioral layer}: specifies with internal actions of a process and communication
performs as seen from the process’ point of view.
\item \textit{Service layer}: it deals with underlying
architectural instructions, states, service instances and correlation sets.
\item \textit{Network layer}: is in charge of connecting and interacting of communicating services.
\end{itemize}

At the current stage of research we have formalized the behavior level of Jolie. We present our results and the detailed description of the behavioral level in the following subsection.

\subsection{Syntax of the behavioral layer}
The most important type of statements in behavioral level regards performing communications and handling data.

\subsubsection{Communications}

There are two types of communication statements in Jolie: input and output statements, both can be uni-directional (one-way operations) and bi-directional (request-response operations). In case of output statements, we use the notion of an output port name (location) which is necessary for binding the communicated data to it. The data is made use by communication statements as variable paths and expressions described below.
\subsubsection{Handling data}
Communication messages in Jolie are represented by means of variable paths structured as a tree. For example:

\begin{center}
  amount = 12\\
  amount.fruit.apple = 2\\
  amount.fruit.description = "Apple"\\
\end{center}

To simplify further operations with variables, we propose their enumeration. Let $ J $ be a Jolie program,
$ V = vars(J) $ -- variables in $ J $, then $ V_i = i$ where $i \in \mathbb{N} $.
Then the example above will look like:

\begin{center}
  0 = 12\\
  1 = 2\\
  2 = "Apple"\\
\end{center}

After this simplification the type of variables can be defined. The type of natural numbers
is located in standard library of Agda~\cite{agdastdlib}.

\begin{hscode}\SaveRestoreHook
\column{B}{@{}>{\hspre}l<{\hspost}@{}}%
\column{E}{@{}>{\hspre}l<{\hspost}@{}}%
\>[B]{}\Conid{Variable}\;\mathbin{:}\;\Conid{Set}{}\<[E]%
\\
\>[B]{}\Conid{Variable}\;\mathrel{=}\;\Conid{ℕ}{}\<[E]%
\ColumnHook
\end{hscode}\resethooks

Complete syntax of behavioral layer can be found in~\cite{nielsen2013type}.
We do not need to consider expressions' structure to prove desired theorems
therefore type Expr is left empty.

\begin{hscode}\SaveRestoreHook
\column{B}{@{}>{\hspre}l<{\hspost}@{}}%
\column{E}{@{}>{\hspre}l<{\hspost}@{}}%
\>[B]{}\Keyword{data}\;\Conid{Expr}\;\mathbin{:}\;\Conid{Set}\;\Keyword{where}{}\<[E]%
\ColumnHook
\end{hscode}\resethooks

Operation names, channel names and locations are represented by strings.

\begin{hscode}\SaveRestoreHook
\column{B}{@{}>{\hspre}l<{\hspost}@{}}%
\column{E}{@{}>{\hspre}l<{\hspost}@{}}%
\>[B]{}\Conid{Operation}\;\Conid{Location}\;\Conid{Channel}\;\mathbin{:}\;\Conid{Set}{}\<[E]%
\\
\>[B]{}\Conid{Operation}\;\mathrel{=}\;\Conid{String}{}\<[E]%
\\
\>[B]{}\Conid{Location}\;\mathrel{=}\;\Conid{String}{}\<[E]%
\\
\>[B]{}\Conid{Channel}\;\mathrel{=}\;\Conid{String}{}\<[E]%
\ColumnHook
\end{hscode}\resethooks

\if{False}
\begin{hscode}\SaveRestoreHook
\column{B}{@{}>{\hspre}l<{\hspost}@{}}%
\column{E}{@{}>{\hspre}l<{\hspost}@{}}%
\>[B]{}\Keyword{data}\;\Conid{Eta}\;\mathbin{:}\;\Conid{Set}{}\<[E]%
\\
\>[B]{}\Keyword{data}\;\Conid{Eta\char94 }\;\mathbin{:}\;\Conid{Set}{}\<[E]%
\ColumnHook
\end{hscode}\resethooks
\fi

The behavioural layer has both ordinary control--flow statements ('if-then-else', 'while', 'assign')
and special statements to control parallelism and communication ('inputchoice', 'parallel', 'input', 'output', etc).

\begin{hscode}\SaveRestoreHook
\column{B}{@{}>{\hspre}l<{\hspost}@{}}%
\column{3}{@{}>{\hspre}l<{\hspost}@{}}%
\column{18}{@{}>{\hspre}l<{\hspost}@{}}%
\column{E}{@{}>{\hspre}l<{\hspost}@{}}%
\>[B]{}\Keyword{data}\;\Conid{Behaviour}\;\mathbin{:}\;\Conid{Set}\;\Keyword{where}{}\<[E]%
\\
\>[B]{}\hsindent{3}{}\<[3]%
\>[3]{}\Varid{if\char95 then\char95 else\char95 }\;\mathbin{:}\;\Conid{Expr}\;\Varid{→}\;\Conid{Behaviour}\;\Varid{→}\;\Conid{Behaviour}\;\Varid{→}\;\Conid{Behaviour}{}\<[E]%
\\
\>[B]{}\hsindent{3}{}\<[3]%
\>[3]{}\Varid{while[\char95 ]\char95 }\;\mathbin{:}\;\Conid{Expr}\;\Varid{→}\;\Conid{Behaviour}\;\Varid{→}\;\Conid{Behaviour}{}\<[E]%
\\[\blanklineskip]%
\>[B]{}\hsindent{3}{}\<[3]%
\>[3]{}\mbox{\onelinecomment  Sequence}{}\<[E]%
\\
\>[B]{}\hsindent{3}{}\<[3]%
\>[3]{}\Varid{\char95 ⇒\char95 }\;\mathbin{:}\;\Conid{Behaviour}\;\Varid{→}\;\Conid{Behaviour}\;\Varid{→}\;\Conid{Behaviour}{}\<[E]%
\\[\blanklineskip]%
\>[B]{}\hsindent{3}{}\<[3]%
\>[3]{}\mbox{\onelinecomment  Parallel}{}\<[E]%
\\
\>[B]{}\hsindent{3}{}\<[3]%
\>[3]{}\Varid{\char95 ∥\char95 }\;\mathbin{:}\;\Conid{Behaviour}\;\Varid{→}\;\Conid{Behaviour}\;\Varid{→}\;\Conid{Behaviour}{}\<[E]%
\\[\blanklineskip]%
\>[B]{}\hsindent{3}{}\<[3]%
\>[3]{}\mbox{\onelinecomment  Assign}{}\<[E]%
\\
\>[B]{}\hsindent{3}{}\<[3]%
\>[3]{}\Varid{\char95 ≃\char95 }\;\mathbin{:}\;\Conid{Variable}\;\Varid{→}\;\Conid{Expr}\;\Varid{→}\;\Conid{Behaviour}{}\<[E]%
\\[\blanklineskip]%
\>[B]{}\hsindent{3}{}\<[3]%
\>[3]{}\Varid{nil}\;\mathbin{:}\;\Conid{Behaviour}{}\<[E]%
\\[\blanklineskip]%
\>[B]{}\hsindent{3}{}\<[3]%
\>[3]{}\mbox{\onelinecomment  [eta\_1]{B\_1}⋯[eta\_a]{B\_a}}{}\<[E]%
\\
\>[B]{}\hsindent{3}{}\<[3]%
\>[3]{}\Varid{inputchoice}\;\mathbin{:}\;\Conid{List}\;(\Conid{Eta}\;\Varid{×}\;\Conid{Behaviour})\;\Varid{→}\;\Conid{Behaviour}{}\<[E]%
\\[\blanklineskip]%
\>[B]{}\hsindent{3}{}\<[3]%
\>[3]{}\Varid{wait}\;\mathbin{:}\;\Conid{Channel}\;\Varid{→}\;\Conid{Operation}\;\Varid{→}\;\Conid{Location}\;\Varid{→}\;\Conid{Variable}\;\Varid{→}\;\Conid{Behaviour}{}\<[E]%
\\
\>[B]{}\hsindent{3}{}\<[3]%
\>[3]{}\Varid{exec}\;\mathbin{:}\;\Conid{Channel}\;\Varid{→}\;\Conid{Operation}\;\Varid{→}\;\Conid{Variable}\;\Varid{→}\;\Conid{Behaviour}\;\Varid{→}\;\Conid{Behaviour}{}\<[E]%
\\[\blanklineskip]%
\>[B]{}\hsindent{3}{}\<[3]%
\>[3]{}\Varid{input}\;\mathbin{:}\;\Conid{Eta}\;\Varid{→}\;\Conid{Behaviour}{}\<[E]%
\\
\>[B]{}\hsindent{3}{}\<[3]%
\>[3]{}\Varid{output}\;\mathbin{:}\;\Conid{Eta\char94 }\;{}\<[18]%
\>[18]{}\Varid{→}\;\Conid{Behaviour}{}\<[E]%
\ColumnHook
\end{hscode}\resethooks

The following two types are called communication ports. They define how communications with other services are performed. There are two kinds of ports:

\begin{itemize}
\item \textit{Input ports}: they deal with exposing input operations to other services.
\item \textit{Output ports}: they define how to invoke a set of operations of other services.
\end{itemize}

\begin{hscode}\SaveRestoreHook
\column{B}{@{}>{\hspre}l<{\hspost}@{}}%
\column{3}{@{}>{\hspre}l<{\hspost}@{}}%
\column{E}{@{}>{\hspre}l<{\hspost}@{}}%
\>[B]{}\mbox{\onelinecomment  Input}{}\<[E]%
\\
\>[B]{}\Keyword{data}\;\Conid{Eta}\;\Keyword{where}{}\<[E]%
\\
\>[B]{}\hsindent{3}{}\<[3]%
\>[3]{}\mbox{\onelinecomment  o(x) -- One-way}{}\<[E]%
\\
\>[B]{}\hsindent{3}{}\<[3]%
\>[3]{}\Varid{\char95 [\char95 ]}\;\mathbin{:}\;\Conid{Operation}\;\Varid{→}\;\Conid{Variable}\;\Varid{→}\;\Conid{Eta}{}\<[E]%
\\[\blanklineskip]%
\>[B]{}\hsindent{3}{}\<[3]%
\>[3]{}\mbox{\onelinecomment  o(x)(x'){B} -- Request-response}{}\<[E]%
\\
\>[B]{}\hsindent{3}{}\<[3]%
\>[3]{}\Varid{\char95 [\char95 ][\char95 ]\char95 }\;\mathbin{:}\;\Conid{Operation}\;\Varid{→}\;\Conid{Variable}\;\Varid{→}\;\Conid{Variable}\;\Varid{→}\;\Conid{Behaviour}\;\Varid{→}\;\Conid{Eta}{}\<[E]%
\\[\blanklineskip]%
\>[B]{}\mbox{\onelinecomment  Output}{}\<[E]%
\\
\>[B]{}\Keyword{data}\;\Conid{Eta\char94 }\;\Keyword{where}{}\<[E]%
\\
\>[B]{}\hsindent{3}{}\<[3]%
\>[3]{}\mbox{\onelinecomment  o at l(e) -- Notification}{}\<[E]%
\\
\>[B]{}\hsindent{3}{}\<[3]%
\>[3]{}\Varid{\char95 at\char95 [\char95 ]}\;\mathbin{:}\;\Conid{Operation}\;\Varid{→}\;\Conid{Location}\;\Varid{→}\;\Conid{Expr}\;\Varid{→}\;\Conid{Eta\char94 }{}\<[E]%
\\[\blanklineskip]%
\>[B]{}\hsindent{3}{}\<[3]%
\>[3]{}\mbox{\onelinecomment  o at l(e)(x) -- Solicit-response}{}\<[E]%
\\
\>[B]{}\hsindent{3}{}\<[3]%
\>[3]{}\Varid{\char95 at\char95 [\char95 ][\char95 ]}\;\mathbin{:}\;\Conid{Operation}\;\Varid{→}\;\Conid{Location}\;\Varid{→}\;\Conid{Expr}\;\Varid{→}\;\Conid{Variable}\;\Varid{→}\;\Conid{Eta\char94 }{}\<[E]%
\ColumnHook
\end{hscode}\resethooks

\section{Jolie type system}\label{sec:typesys}
Jolie type system consists of
commonly-used native types as \jkeyw{int}, \jkeyw{double}, \jkeyw{long}, \jkeyw{boolean} and \jkeyw{string}, Jolie also has the following types:

\begin{itemize}
\item \jkeyw{raw}, representing raw data streams as byte arrays.
\item \jkeyw{void}, indicating no value.
\item \jkeyw{any}, as a placeholder for any native types.
\end{itemize}

In order to be able to do type checking of a Jolie program, we need to provide implementation of types and typing rules in Agda.

\subsection{Type declaration}
We express main Jolie native types (excluding any) in the following way:

\begin{hscode}\SaveRestoreHook
\column{B}{@{}>{\hspre}l<{\hspost}@{}}%
\column{3}{@{}>{\hspre}l<{\hspost}@{}}%
\column{E}{@{}>{\hspre}l<{\hspost}@{}}%
\>[B]{}\Keyword{data}\;\Conid{Type}\;\mathbin{:}\;\Conid{Set}\;\Keyword{where}{}\<[E]%
\\
\>[B]{}\hsindent{3}{}\<[3]%
\>[3]{}\Varid{bool}\;\Varid{int}\;\Varid{double}\;\Varid{long}\;\Varid{string}\;\Varid{raw}\;\Varid{void}\;\mathbin{:}\;\Conid{Type}{}\<[E]%
\ColumnHook
\end{hscode}\resethooks

Usually, the context of a program is a list of variables, but to service all three layers (comprising communication of services) there is a special type called TypeDecl. It has five constructors: the first two (unidirectional and bidirectional) are for output communication. The left part of such bindings consists of an operation name and a location of a hosting service. The next two are for input communication and the last one is for variables.

\begin{hscode}\SaveRestoreHook
\column{B}{@{}>{\hspre}l<{\hspost}@{}}%
\column{3}{@{}>{\hspre}l<{\hspost}@{}}%
\column{E}{@{}>{\hspre}l<{\hspost}@{}}%
\>[B]{}\Keyword{data}\;\Conid{TypeDecl}\;\mathbin{:}\;\Conid{Set}\;\Keyword{where}{}\<[E]%
\\
\>[B]{}\hsindent{3}{}\<[3]%
\>[3]{}\mbox{\onelinecomment  o at l : <T>}{}\<[E]%
\\
\>[B]{}\hsindent{3}{}\<[3]%
\>[3]{}\Varid{\char95 at\char95 ∶<\char95 >}\;\mathbin{:}\;\Conid{Operation}\;\Varid{→}\;\Conid{Location}\;\Varid{→}\;\Conid{Type}\;\Varid{→}\;\Conid{TypeDecl}{}\<[E]%
\\[\blanklineskip]%
\>[B]{}\hsindent{3}{}\<[3]%
\>[3]{}\mbox{\onelinecomment  o at l : <T, T>}{}\<[E]%
\\
\>[B]{}\hsindent{3}{}\<[3]%
\>[3]{}\Varid{\char95 at\char95 ∶<\char95 ,\char95 >}\;\mathbin{:}\;\Conid{Operation}\;\Varid{→}\;\Conid{Location}\;\Varid{→}\;\Conid{Type}\;\Varid{→}\;\Conid{Type}\;\Varid{→}\;\Conid{TypeDecl}{}\<[E]%
\\[\blanklineskip]%
\>[B]{}\hsindent{3}{}\<[3]%
\>[3]{}\mbox{\onelinecomment  o : <T>}{}\<[E]%
\\
\>[B]{}\hsindent{3}{}\<[3]%
\>[3]{}\Varid{\char95 ∶<\char95 >}\;\mathbin{:}\;\Conid{Operation}\;\Varid{→}\;\Conid{Type}\;\Varid{→}\;\Conid{TypeDecl}{}\<[E]%
\\[\blanklineskip]%
\>[B]{}\hsindent{3}{}\<[3]%
\>[3]{}\mbox{\onelinecomment  o : <T, T>}{}\<[E]%
\\
\>[B]{}\hsindent{3}{}\<[3]%
\>[3]{}\Varid{\char95 ∶<\char95 ,\char95 >}\;\mathbin{:}\;\Conid{Operation}\;\Varid{→}\;\Conid{Type}\;\Varid{→}\;\Conid{Type}\;\Varid{→}\;\Conid{TypeDecl}{}\<[E]%
\\[\blanklineskip]%
\>[B]{}\hsindent{3}{}\<[3]%
\>[3]{}\mbox{\onelinecomment  x : T}{}\<[E]%
\\
\>[B]{}\hsindent{3}{}\<[3]%
\>[3]{}\Varid{\char95 ∶\char95 }\;\mathbin{:}\;\Conid{Variable}\;\Varid{→}\;\Conid{Type}\;\Varid{→}\;\Conid{TypeDecl}{}\<[E]%
\ColumnHook
\end{hscode}\resethooks

Therefore, the type of context is a vector of TypeDecl.

\begin{hscode}\SaveRestoreHook
\column{B}{@{}>{\hspre}l<{\hspost}@{}}%
\column{E}{@{}>{\hspre}l<{\hspost}@{}}%
\>[B]{}\Conid{Ctx}\;\mathbin{:}\;\Conid{ℕ}\;\Varid{→}\;\Conid{Set}{}\<[E]%
\\
\>[B]{}\Conid{Ctx}\;\mathrel{=}\;\Conid{Vec}\;\Conid{TypeDecl}{}\<[E]%
\ColumnHook
\end{hscode}\resethooks

Although the type of context is defined, it's not enough, because programs in Jolie can be parallel. We define one more type called Context to cover such situations. It has only two constructors: the first one just takes Ctx\ n and the second one consists of two elements of itself.

\begin{hscode}\SaveRestoreHook
\column{B}{@{}>{\hspre}l<{\hspost}@{}}%
\column{3}{@{}>{\hspre}l<{\hspost}@{}}%
\column{E}{@{}>{\hspre}l<{\hspost}@{}}%
\>[B]{}\Keyword{data}\;\Conid{Context}\;\mathbin{:}\;\Conid{Set}\;\Keyword{where}{}\<[E]%
\\
\>[B]{}\hsindent{3}{}\<[3]%
\>[3]{}\Varid{⋆}\;\mathbin{:}\;\Varid{∀}\;\{\mskip1.5mu \Varid{n}\mskip1.5mu\}\;\Varid{→}\;\Conid{Ctx}\;\Varid{n}\;\Varid{→}\;\Conid{Context}{}\<[E]%
\\
\>[B]{}\hsindent{3}{}\<[3]%
\>[3]{}\Varid{\&}\;\mathbin{:}\;\Conid{Context}\;\Varid{→}\;\Conid{Context}\;\Varid{→}\;\Conid{Context}{}\<[E]%
\ColumnHook
\end{hscode}\resethooks

The type of context is not a vector anymore, so we need to define such type that will express the fact of presence of TypeDecl in Context.

\begin{hscode}\SaveRestoreHook
\column{B}{@{}>{\hspre}l<{\hspost}@{}}%
\column{3}{@{}>{\hspre}l<{\hspost}@{}}%
\column{10}{@{}>{\hspre}l<{\hspost}@{}}%
\column{13}{@{}>{\hspre}l<{\hspost}@{}}%
\column{15}{@{}>{\hspre}l<{\hspost}@{}}%
\column{16}{@{}>{\hspre}l<{\hspost}@{}}%
\column{17}{@{}>{\hspre}l<{\hspost}@{}}%
\column{18}{@{}>{\hspre}l<{\hspost}@{}}%
\column{24}{@{}>{\hspre}l<{\hspost}@{}}%
\column{E}{@{}>{\hspre}l<{\hspost}@{}}%
\>[B]{}\Keyword{infix}\;\Varid{4}\;\Varid{\char95 ∈\char95 }{}\<[E]%
\\
\>[B]{}\Keyword{data}\;\Varid{\char95 ∈\char95 }\;\mathbin{:}\;\Conid{TypeDecl}\;\Varid{→}\;\Conid{Context}\;\Varid{→}\;\Conid{Set}\;\Keyword{where}{}\<[E]%
\\
\>[B]{}\hsindent{3}{}\<[3]%
\>[3]{}\Varid{here-⋆}\;\mathbin{:}\;\Varid{∀}\;\{\mskip1.5mu \Varid{n}\mskip1.5mu\}\;\{\mskip1.5mu \Varid{x}\mskip1.5mu\}\;\{\mskip1.5mu \Varid{xs}\;\mathbin{:}\;\Conid{Ctx}\;\Varid{n}\mskip1.5mu\}\;{}\<[E]%
\\
\>[3]{}\hsindent{7}{}\<[10]%
\>[10]{}\Varid{→}\;\Varid{x}\;\Varid{∈}\;\Varid{⋆}\;(\Varid{x}\;\Varid{∷}\;\Varid{xs}){}\<[E]%
\\[\blanklineskip]%
\>[B]{}\hsindent{3}{}\<[3]%
\>[3]{}\Varid{there-⋆}\;\mathbin{:}\;\Varid{∀}\;\{\mskip1.5mu \Varid{n}\mskip1.5mu\}\;\{\mskip1.5mu \Varid{x}\;\Varid{y}\mskip1.5mu\}\;\{\mskip1.5mu \Varid{xs}\;\mathbin{:}\;\Conid{Ctx}\;\Varid{n}\mskip1.5mu\}\;{}\<[E]%
\\
\>[3]{}\hsindent{10}{}\<[13]%
\>[13]{}(\Varid{x∈xs}\;\mathbin{:}\;\Varid{x}\;\Varid{∈}\;\Varid{⋆}\;\Varid{xs})\;{}\<[E]%
\\
\>[3]{}\hsindent{10}{}\<[13]%
\>[13]{}\Varid{→}\;\Varid{x}\;\Varid{∈}\;\Varid{⋆}\;(\Varid{y}\;\Varid{∷}\;\Varid{xs}){}\<[E]%
\\[\blanklineskip]%
\>[B]{}\hsindent{3}{}\<[3]%
\>[3]{}\Varid{here-left-\&}\;\mathbin{:}\;\Varid{∀}\;\{\mskip1.5mu \Varid{n}\;\Varid{m}\mskip1.5mu\}\;\{\mskip1.5mu \Varid{x}\mskip1.5mu\}\;\{\mskip1.5mu \Varid{xs}\;\mathbin{:}\;\Conid{Ctx}\;\Varid{n}\mskip1.5mu\}\;\{\mskip1.5mu \Varid{ys}\;\mathbin{:}\;\Conid{Ctx}\;\Varid{m}\mskip1.5mu\}\;{}\<[E]%
\\
\>[3]{}\hsindent{12}{}\<[15]%
\>[15]{}\Varid{→}\;\Varid{x}\;\Varid{∈}\;\Varid{\&}\;{}\<[24]%
\>[24]{}(\Varid{⋆}\;(\Varid{x}\;\Varid{∷}\;\Varid{xs}))\;(\Varid{⋆}\;\Varid{ys}){}\<[E]%
\\[\blanklineskip]%
\>[B]{}\hsindent{3}{}\<[3]%
\>[3]{}\Varid{here-right-\&}\;\mathbin{:}\;\Varid{∀}\;\{\mskip1.5mu \Varid{n}\;\Varid{m}\mskip1.5mu\}\;\{\mskip1.5mu \Varid{x}\mskip1.5mu\}\;\{\mskip1.5mu \Varid{xs}\;\mathbin{:}\;\Conid{Ctx}\;\Varid{n}\mskip1.5mu\}\;\{\mskip1.5mu \Varid{ys}\;\mathbin{:}\;\Conid{Ctx}\;\Varid{m}\mskip1.5mu\}\;{}\<[E]%
\\
\>[3]{}\hsindent{13}{}\<[16]%
\>[16]{}\Varid{→}\;\Varid{x}\;\Varid{∈}\;\Varid{\&}\;(\Varid{⋆}\;\Varid{xs})\;(\Varid{⋆}\;(\Varid{x}\;\Varid{∷}\;\Varid{ys})){}\<[E]%
\\[\blanklineskip]%
\>[B]{}\hsindent{3}{}\<[3]%
\>[3]{}\Varid{there-left-\&}\;\mathbin{:}\;\Varid{∀}\;\{\mskip1.5mu \Varid{n}\;\Varid{m}\mskip1.5mu\}\;\{\mskip1.5mu \Varid{x}\mskip1.5mu\}\;\{\mskip1.5mu \Varid{xs}\;\mathbin{:}\;\Conid{Ctx}\;\Varid{n}\mskip1.5mu\}\;\{\mskip1.5mu \Varid{ys}\;\mathbin{:}\;\Conid{Ctx}\;\Varid{m}\mskip1.5mu\}\;{}\<[E]%
\\
\>[3]{}\hsindent{15}{}\<[18]%
\>[18]{}(\Varid{x∈xs}\;\mathbin{:}\;\Varid{x}\;\Varid{∈}\;\Varid{\&}\;(\Varid{⋆}\;\Varid{xs})\;(\Varid{⋆}\;\Varid{ys}))\;{}\<[E]%
\\
\>[3]{}\hsindent{13}{}\<[16]%
\>[16]{}\Varid{→}\;\Varid{x}\;\Varid{∈}\;\Varid{\&}\;(\Varid{⋆}\;(\Varid{x}\;\Varid{∷}\;\Varid{xs}))\;(\Varid{⋆}\;\Varid{ys}){}\<[E]%
\\[\blanklineskip]%
\>[B]{}\hsindent{3}{}\<[3]%
\>[3]{}\Varid{there-right-\&}\;\mathbin{:}\;\Varid{∀}\;\{\mskip1.5mu \Varid{n}\;\Varid{m}\mskip1.5mu\}\;\{\mskip1.5mu \Varid{x}\mskip1.5mu\}\;\{\mskip1.5mu \Varid{xs}\;\mathbin{:}\;\Conid{Ctx}\;\Varid{n}\mskip1.5mu\}\;\{\mskip1.5mu \Varid{ys}\;\mathbin{:}\;\Conid{Ctx}\;\Varid{m}\mskip1.5mu\}\;{}\<[E]%
\\
\>[3]{}\hsindent{14}{}\<[17]%
\>[17]{}(\Varid{x∈xs}\;\mathbin{:}\;\Varid{x}\;\Varid{∈}\;\Varid{\&}\;(\Varid{⋆}\;\Varid{xs})\;(\Varid{⋆}\;\Varid{ys}))\;{}\<[E]%
\\
\>[3]{}\hsindent{14}{}\<[17]%
\>[17]{}\Varid{→}\;\Varid{x}\;\Varid{∈}\;\Varid{\&}\;(\Varid{⋆}\;\Varid{xs})\;(\Varid{⋆}\;(\Varid{x}\;\Varid{∷}\;\Varid{ys})){}\<[E]%
\ColumnHook
\end{hscode}\resethooks

Since we don't care about expressions at all, we introduce the empty type of a correctly typed expression with variables from context $ \Gamma $.

\begin{hscode}\SaveRestoreHook
\column{B}{@{}>{\hspre}l<{\hspost}@{}}%
\column{E}{@{}>{\hspre}l<{\hspost}@{}}%
\>[B]{}\Keyword{data}\;\Varid{\char95 ⊢e\char95 ∶\char95 }\;(\Conid{Γ}\;\mathbin{:}\;\Conid{Context})\;\mathbin{:}\;\Conid{Expr}\;\Varid{→}\;\Conid{Type}\;\Varid{→}\;\Conid{Set}\;\Keyword{where}{}\<[E]%
\ColumnHook
\end{hscode}\resethooks

\subsection{Typing rules}
Finally, we can present the subset of the typing rules of the behavioural layer. The first constructor is for nil behaviour. Since nil does nothing, the contexts before and after are equal. The next two are rules for ordinary behaviours if\_then\_else and while. Finally, the last two are for sequent and parallel statements.

\begin{hscode}\SaveRestoreHook
\column{B}{@{}>{\hspre}l<{\hspost}@{}}%
\column{3}{@{}>{\hspre}l<{\hspost}@{}}%
\column{8}{@{}>{\hspre}l<{\hspost}@{}}%
\column{9}{@{}>{\hspre}l<{\hspost}@{}}%
\column{11}{@{}>{\hspre}l<{\hspost}@{}}%
\column{E}{@{}>{\hspre}l<{\hspost}@{}}%
\>[B]{}\Keyword{data}\;\Varid{\char95 ⊢B\char95 }\triangleright\;\mathbin{:}\;\Conid{Context}\;\Varid{→}\;\Conid{Behaviour}\;\Varid{→}\;\Conid{Context}\;\Varid{→}\;\Conid{Set}\;\Keyword{where}{}\<[E]%
\\
\>[B]{}\hsindent{3}{}\<[3]%
\>[3]{}\Varid{t-nil}\;\mathbin{:}\;\{\mskip1.5mu \Conid{Γ}\;\mathbin{:}\;\Conid{Context}\mskip1.5mu\}\;{}\<[E]%
\\
\>[3]{}\hsindent{6}{}\<[9]%
\>[9]{}\Varid{→}\;\Conid{Γ}\;\Varid{⊢B}\;\Varid{nil}\;\triangleright\;\Conid{Γ}{}\<[E]%
\\[\blanklineskip]%
\>[B]{}\hsindent{3}{}\<[3]%
\>[3]{}\Varid{t-if}\;\mathbin{:}\;\{\mskip1.5mu \Conid{Γ}\;\Conid{Γ1}\;\mathbin{:}\;\Conid{Context}\mskip1.5mu\}\;\{\mskip1.5mu \Varid{b1}\;\Varid{b2}\;\mathbin{:}\;\Conid{Behaviour}\mskip1.5mu\}\;\{\mskip1.5mu \Varid{e}\;\mathbin{:}\;\Conid{Expr}\mskip1.5mu\}\;{}\<[E]%
\\
\>[3]{}\hsindent{5}{}\<[8]%
\>[8]{}\Varid{→}\;\Conid{Γ}\;\Varid{⊢e}\;\Varid{e}\;\Varid{∶}\;\Varid{bool}\;{}\<[E]%
\\
\>[3]{}\hsindent{5}{}\<[8]%
\>[8]{}\Varid{→}\;\Conid{Γ}\;\Varid{⊢B}\;\Varid{b1}\;\triangleright\;\Conid{Γ1}\;{}\<[E]%
\\
\>[3]{}\hsindent{5}{}\<[8]%
\>[8]{}\Varid{→}\;\Conid{Γ}\;\Varid{⊢B}\;\Varid{b2}\;\triangleright\;\Conid{Γ1}\;{}\<[E]%
\\
\>[3]{}\hsindent{5}{}\<[8]%
\>[8]{}\Varid{→}\;\Conid{Γ}\;\Varid{⊢B}\;\Varid{if}\;\Varid{e}\;\Varid{then}\;\Varid{b1}\;\Varid{else}\;\Varid{b2}\;\triangleright\;\Conid{Γ1}{}\<[E]%
\\[\blanklineskip]%
\>[B]{}\hsindent{3}{}\<[3]%
\>[3]{}\Varid{t-while}\;\mathbin{:}\;\{\mskip1.5mu \Conid{Γ}\;\mathbin{:}\;\Conid{Context}\mskip1.5mu\}\;\{\mskip1.5mu \Varid{b}\;\mathbin{:}\;\Conid{Behaviour}\mskip1.5mu\}\;\{\mskip1.5mu \Varid{e}\;\mathbin{:}\;\Conid{Expr}\mskip1.5mu\}\;{}\<[E]%
\\
\>[3]{}\hsindent{8}{}\<[11]%
\>[11]{}\Varid{→}\;\Conid{Γ}\;\Varid{⊢e}\;\Varid{e}\;\Varid{∶}\;\Varid{bool}\;{}\<[E]%
\\
\>[3]{}\hsindent{8}{}\<[11]%
\>[11]{}\Varid{→}\;\Conid{Γ}\;\Varid{⊢B}\;\Varid{b}\;\triangleright\;\Conid{Γ}\;{}\<[E]%
\\
\>[3]{}\hsindent{8}{}\<[11]%
\>[11]{}\Varid{→}\;\Conid{Γ}\;\Varid{⊢B}\;\Varid{while[}\;\Varid{e}\;\mskip1.5mu]\;\Varid{b}\;\triangleright\;\Conid{Γ}{}\<[E]%
\\[\blanklineskip]%
\>[B]{}\hsindent{3}{}\<[3]%
\>[3]{}\Varid{t-seq}\;\mathbin{:}\;\{\mskip1.5mu \Conid{Γ}\;\Conid{Γ1}\;\Conid{Γ2}\;\mathbin{:}\;\Conid{Context}\mskip1.5mu\}\;\{\mskip1.5mu \Varid{b1}\;\Varid{b2}\;\mathbin{:}\;\Conid{Behaviour}\mskip1.5mu\}\;{}\<[E]%
\\
\>[3]{}\hsindent{6}{}\<[9]%
\>[9]{}\Varid{→}\;\Conid{Γ}\;\Varid{⊢B}\;\Varid{b1}\;\triangleright\;\Conid{Γ1}\;{}\<[E]%
\\
\>[3]{}\hsindent{6}{}\<[9]%
\>[9]{}\Varid{→}\;\Conid{Γ1}\;\Varid{⊢B}\;\Varid{b2}\;\triangleright\;\Conid{Γ2}\;{}\<[E]%
\\
\>[3]{}\hsindent{6}{}\<[9]%
\>[9]{}\Varid{→}\;\Conid{Γ}\;\Varid{⊢B}\;\Varid{b1}\;\Varid{⇒}\;\Varid{b2}\;\triangleright\;\Conid{Γ2}{}\<[E]%
\\[\blanklineskip]%
\>[B]{}\hsindent{3}{}\<[3]%
\>[3]{}\Varid{t-par}\;\mathbin{:}\;\{\mskip1.5mu \Conid{Γ1}\;\Conid{Γ2}\;\Conid{Γ1'}\;\Conid{Γ2'}\;\mathbin{:}\;\Conid{Context}\mskip1.5mu\}\;\{\mskip1.5mu \Varid{b1}\;\Varid{b2}\;\mathbin{:}\;\Conid{Behaviour}\mskip1.5mu\}\;{}\<[E]%
\\
\>[3]{}\hsindent{6}{}\<[9]%
\>[9]{}\Varid{→}\;\Conid{Γ1}\;\Varid{⊢B}\;\Varid{b1}\;\triangleright\;\Conid{Γ1'}\;{}\<[E]%
\\
\>[3]{}\hsindent{6}{}\<[9]%
\>[9]{}\Varid{→}\;\Conid{Γ2}\;\Varid{⊢B}\;\Varid{b2}\;\triangleright\;\Conid{Γ2'}\;{}\<[E]%
\\
\>[3]{}\hsindent{6}{}\<[9]%
\>[9]{}\Varid{→}\;(\Varid{\&}\;\Conid{Γ1}\;\Conid{Γ2})\;\Varid{⊢B}\;\Varid{b1}\;\Varid{∥}\;\Varid{b2}\;\triangleright\;(\Varid{\&}\;\Conid{Γ1'}\;\Conid{Γ2'}){}\<[E]%
\ColumnHook
\end{hscode}\resethooks

\section{Structural Congruence for Behaviours}\label{sec:strcongr}

According to the Curry-Howard isomorphism~\cite{Sorensen}, types of the programs are propostions and terms are proofs. For example, the type $ A \rightarrow B $ correspond to the implication from $ A $ to $ B $ and such function $ f $ that takes an element of type $ A $ and returns an element of type $ B $ will be a proof of this theorem.

To demonstrate the correctness of the typing rules given above, we will prove the lemma called "Structural Congruence for Behaviours"~\cite{nielsen2013type,engelfriet}:

\begin{center}
\textit{Let} $ \Gamma \vdash B_1 \rhd \Gamma' $ \\
\textit{If} $ B_1 \equiv B_2 $ \\
\textit{then} $ \Gamma \vdash B_2 \rhd \Gamma' $
\end{center}

The proof is the case analysis of all possible $ B_1 $ and $ B_2 $.

\begin{itemize}

\item \textit{Case} $ B_1 \equiv B_2 $

\begin{hscode}\SaveRestoreHook
\column{B}{@{}>{\hspre}l<{\hspost}@{}}%
\column{19}{@{}>{\hspre}l<{\hspost}@{}}%
\column{E}{@{}>{\hspre}l<{\hspost}@{}}%
\>[B]{}\Varid{struct-cong-b₁≡b₂}\;\mathbin{:}\;\{\mskip1.5mu \Conid{Γ}\;\Conid{Γ₁}\;\mathbin{:}\;\Conid{Context}\mskip1.5mu\}\;\{\mskip1.5mu \Varid{b₁}\;\Varid{b₂}\;\mathbin{:}\;\Conid{Behaviour}\mskip1.5mu\}\;{}\<[E]%
\\
\>[B]{}\hsindent{19}{}\<[19]%
\>[19]{}\Varid{→}\;\Conid{Γ}\;\Varid{⊢B}\;\Varid{b₁}\;\triangleright\;\Conid{Γ₁}\;{}\<[E]%
\\
\>[B]{}\hsindent{19}{}\<[19]%
\>[19]{}\Varid{→}\;\Varid{b₁}\;\Varid{≡}\;\Varid{b₂}\;{}\<[E]%
\\
\>[B]{}\hsindent{19}{}\<[19]%
\>[19]{}\Varid{→}\;\Conid{Γ}\;\Varid{⊢B}\;\Varid{b₂}\;\triangleright\;\Conid{Γ₁}{}\<[E]%
\\
\>[B]{}\Varid{struct-cong-b₁≡b₂}\;\Varid{t}\;\Varid{refl}\;\mathrel{=}\;\Varid{t}{}\<[E]%
\ColumnHook
\end{hscode}\resethooks

\item \textit{Case} $ 0; B \equiv B $

\begin{hscode}\SaveRestoreHook
\column{B}{@{}>{\hspre}l<{\hspost}@{}}%
\column{21}{@{}>{\hspre}l<{\hspost}@{}}%
\column{E}{@{}>{\hspre}l<{\hspost}@{}}%
\>[B]{}\Varid{struct-cong-nil∶b→b}\;\mathbin{:}\;\{\mskip1.5mu \Conid{Γ}\;\Conid{Γ₁}\;\mathbin{:}\;\Conid{Context}\mskip1.5mu\}\;\{\mskip1.5mu \Varid{b}\;\mathbin{:}\;\Conid{Behaviour}\mskip1.5mu\}\;{}\<[E]%
\\
\>[B]{}\hsindent{21}{}\<[21]%
\>[21]{}\Varid{→}\;\Conid{Γ}\;\Varid{⊢B}\;\Varid{nil}\;\Varid{⇒}\;\Varid{b}\;\triangleright\;\Conid{Γ₁}\;{}\<[E]%
\\
\>[B]{}\hsindent{21}{}\<[21]%
\>[21]{}\Varid{→}\;\Conid{Γ}\;\Varid{⊢B}\;\Varid{b}\;\triangleright\;\Conid{Γ₁}{}\<[E]%
\\
\>[B]{}\Varid{struct-cong-nil∶b→b}\;(\Varid{t-seq}\;\Varid{t-nil}\;\Varid{x})\;\mathrel{=}\;\Varid{x}{}\<[E]%
\ColumnHook
\end{hscode}\resethooks

\item \textit{Case} $ B \equiv 0 ; B $

\begin{hscode}\SaveRestoreHook
\column{B}{@{}>{\hspre}l<{\hspost}@{}}%
\column{21}{@{}>{\hspre}l<{\hspost}@{}}%
\column{E}{@{}>{\hspre}l<{\hspost}@{}}%
\>[B]{}\Varid{struct-cong-b→nil∶b}\;\mathbin{:}\;\{\mskip1.5mu \Conid{Γ}\;\Conid{Γ₁}\;\mathbin{:}\;\Conid{Context}\mskip1.5mu\}\;\{\mskip1.5mu \Varid{b}\;\mathbin{:}\;\Conid{Behaviour}\mskip1.5mu\}\;{}\<[E]%
\\
\>[B]{}\hsindent{21}{}\<[21]%
\>[21]{}\Varid{→}\;\Conid{Γ}\;\Varid{⊢B}\;\Varid{b}\;\triangleright\;\Conid{Γ₁}\;{}\<[E]%
\\
\>[B]{}\hsindent{21}{}\<[21]%
\>[21]{}\Varid{→}\;\Conid{Γ}\;\Varid{⊢B}\;\Varid{nil}\;\Varid{⇒}\;\Varid{b}\;\triangleright\;\Conid{Γ₁}{}\<[E]%
\\
\>[B]{}\Varid{struct-cong-b→nil∶b}\;\Varid{x}\;\mathrel{=}\;\Varid{t-seq}\;\Varid{t-nil}\;\Varid{x}{}\<[E]%
\ColumnHook
\end{hscode}\resethooks

\item \textit{Case} $ B \parallel 0 \equiv B $
\begin{hscode}\SaveRestoreHook
\column{B}{@{}>{\hspre}l<{\hspost}@{}}%
\column{21}{@{}>{\hspre}l<{\hspost}@{}}%
\column{E}{@{}>{\hspre}l<{\hspost}@{}}%
\>[B]{}\Varid{struct-cong-b∥nil→b}\;\mathbin{:}\;\{\mskip1.5mu \Conid{Γ₁}\;\Conid{Γ₂}\;\Conid{Γ₁'}\;\Conid{Γ₂'}\;\mathbin{:}\;\Conid{Context}\mskip1.5mu\}\;\{\mskip1.5mu \Varid{b}\;\mathbin{:}\;\Conid{Behaviour}\mskip1.5mu\}\;{}\<[E]%
\\
\>[B]{}\hsindent{21}{}\<[21]%
\>[21]{}\Varid{→}\;\Varid{\&}\;\Conid{Γ₁}\;\Conid{Γ₂}\;\Varid{⊢B}\;(\Varid{b}\;\Varid{∥}\;\Varid{nil})\;\triangleright\;\Varid{\&}\;\Conid{Γ₁'}\;\Conid{Γ₂'}\;{}\<[E]%
\\
\>[B]{}\hsindent{21}{}\<[21]%
\>[21]{}\Varid{→}\;\Conid{Γ₁}\;\Varid{⊢B}\;\Varid{b}\;\triangleright\;\Conid{Γ₁'}{}\<[E]%
\\
\>[B]{}\Varid{struct-cong-b∥nil→b}\;(\Varid{t-par}\;\Varid{x}\;\anonymous )\;\mathrel{=}\;\Varid{x}{}\<[E]%
\ColumnHook
\end{hscode}\resethooks

\item \textit{Case} $ B \equiv B \parallel 0 $

\begin{hscode}\SaveRestoreHook
\column{B}{@{}>{\hspre}l<{\hspost}@{}}%
\column{21}{@{}>{\hspre}l<{\hspost}@{}}%
\column{E}{@{}>{\hspre}l<{\hspost}@{}}%
\>[B]{}\Varid{struct-cong-b→b∥nil}\;\mathbin{:}\;\{\mskip1.5mu \Conid{Γ₁}\;\Conid{Γ₂}\;\Conid{Γ₃}\;\mathbin{:}\;\Conid{Context}\mskip1.5mu\}\;\{\mskip1.5mu \Varid{b}\;\mathbin{:}\;\Conid{Behaviour}\mskip1.5mu\}\;{}\<[E]%
\\
\>[B]{}\hsindent{21}{}\<[21]%
\>[21]{}\Varid{→}\;\Conid{Γ₁}\;\Varid{⊢B}\;\Varid{b}\;\triangleright\;\Conid{Γ₂}\;{}\<[E]%
\\
\>[B]{}\hsindent{21}{}\<[21]%
\>[21]{}\Varid{→}\;\Varid{\&}\;\Conid{Γ₁}\;\Conid{Γ₃}\;\Varid{⊢B}\;(\Varid{b}\;\Varid{∥}\;\Varid{nil})\;\triangleright\;\Varid{\&}\;\Conid{Γ₂}\;\Conid{Γ₃}{}\<[E]%
\\
\>[B]{}\Varid{struct-cong-b→b∥nil}\;\Varid{x}\;\mathrel{=}\;\Varid{t-par}\;\Varid{x}\;\Varid{t-nil}{}\<[E]%
\ColumnHook
\end{hscode}\resethooks

\item \textit{Case} $ B_1 \parallel B_2 \equiv B_2 \parallel B_1 $

\begin{hscode}\SaveRestoreHook
\column{B}{@{}>{\hspre}l<{\hspost}@{}}%
\column{21}{@{}>{\hspre}l<{\hspost}@{}}%
\column{E}{@{}>{\hspre}l<{\hspost}@{}}%
\>[B]{}\Varid{struct-cong-par-comm}\;\mathbin{:}\;\{\mskip1.5mu \Conid{Γ₁}\;\Conid{Γ₂}\;\Conid{Γ₁'}\;\Conid{Γ₂'}\;\mathbin{:}\;\Conid{Context}\mskip1.5mu\}\;\{\mskip1.5mu \Varid{b₁}\;\Varid{b₂}\;\mathbin{:}\;\Conid{Behaviour}\mskip1.5mu\}\;{}\<[E]%
\\
\>[B]{}\hsindent{21}{}\<[21]%
\>[21]{}\Varid{→}\;\Varid{\&}\;\Conid{Γ₁}\;\Conid{Γ₂}\;\Varid{⊢B}\;(\Varid{b₁}\;\Varid{∥}\;\Varid{b₂})\;\triangleright\;\Varid{\&}\;\Conid{Γ₁'}\;\Conid{Γ₂'}\;{}\<[E]%
\\
\>[B]{}\hsindent{21}{}\<[21]%
\>[21]{}\Varid{→}\;\Varid{\&}\;\Conid{Γ₂}\;\Conid{Γ₁}\;\Varid{⊢B}\;(\Varid{b₂}\;\Varid{∥}\;\Varid{b₁})\;\triangleright\;\Varid{\&}\;\Conid{Γ₂'}\;\Conid{Γ₁'}{}\<[E]%
\\
\>[B]{}\Varid{struct-cong-par-comm}\;(\Varid{t-par}\;\Varid{t₁}\;\Varid{t₂})\;\mathrel{=}\;\Varid{t-par}\;\Varid{t₂}\;\Varid{t₁}{}\<[E]%
\ColumnHook
\end{hscode}\resethooks

\item \textit{Case} $ (B_1 \parallel B_2) \parallel B_3 \equiv B_1 \parallel (B_2 \parallel B_3) $

\begin{hscode}\SaveRestoreHook
\column{B}{@{}>{\hspre}l<{\hspost}@{}}%
\column{21}{@{}>{\hspre}l<{\hspost}@{}}%
\column{E}{@{}>{\hspre}l<{\hspost}@{}}%
\>[B]{}\Varid{struct-cong-par-assoc}\;\mathbin{:}\;\{\mskip1.5mu \Conid{Γ₁}\;\Conid{Γ₂}\;\Conid{Γ₃}\;\Conid{Γ₁'}\;\Conid{Γ₂'}\;\Conid{Γ₃'}\;\mathbin{:}\;\Conid{Context}\mskip1.5mu\}\;\{\mskip1.5mu \Varid{b₁}\;\Varid{b₂}\;\Varid{b₃}\;\mathbin{:}\;\Conid{Behaviour}\mskip1.5mu\}\;{}\<[E]%
\\
\>[B]{}\hsindent{21}{}\<[21]%
\>[21]{}\Varid{→}\;\Varid{\&}\;(\Varid{\&}\;\Conid{Γ₁}\;\Conid{Γ₂})\;\Conid{Γ₃}\;\Varid{⊢B}\;(\Varid{b₁}\;\Varid{∥}\;\Varid{b₂})\;\Varid{∥}\;\Varid{b₃}\;\triangleright\;\Varid{\&}\;(\Varid{\&}\;\Conid{Γ₁'}\;\Conid{Γ₂'})\;\Conid{Γ₃'}\;{}\<[E]%
\\
\>[B]{}\hsindent{21}{}\<[21]%
\>[21]{}\Varid{→}\;\Varid{\&}\;\Conid{Γ₁}\;(\Varid{\&}\;\Conid{Γ₂}\;\Conid{Γ₃})\;\Varid{⊢B}\;\Varid{b₁}\;\Varid{∥}\;(\Varid{b₂}\;\Varid{∥}\;\Varid{b₃})\;\triangleright\;\Varid{\&}\;\Conid{Γ₁'}\;(\Varid{\&}\;\Conid{Γ₂'}\;\Conid{Γ₃'}){}\<[E]%
\\
\>[B]{}\Varid{struct-cong-par-assoc}\;(\Varid{t-par}\;(\Varid{t-par}\;\Varid{t1}\;\Varid{t2})\;\Varid{t3})\;\mathrel{=}\;\Varid{t-par}\;\Varid{t1}\;(\Varid{t-par}\;\Varid{t2}\;\Varid{t3}){}\<[E]%
\ColumnHook
\end{hscode}\resethooks

The proof for $ B_1 \parallel (B_2 \parallel B_3) \equiv (B_1 \parallel B_2) \parallel B_3 $ is similar.

\end{itemize}

\section{Conclusions and future work}\label{sec:concl}

In this paper, we presented our approach in creating the static type-checker for Jolie programming language which is currently dynamically type-checked. We developed the formalization for the subset of Jolie by means of Agda proof assistant. We expressed Jolie types and typing rules in order to be able to check the type correspondence of messages which are used for interaction of microservices in Jolie. We have also provided the proof of structural congruence lemma which means the correctness of type checker itself.

However, our current implementation covers only the small subset of Jolie. We have touched only the native types, though the type system of Jolie goes beyond it and includes subtyping, linked types etc.
Another important direction leads us to formalization of the service and communication levels of Jolie. By accomplishing this, we would be able to type-check Jolie program thoroughly and may even think about implementing of verifiable compiler for Jolie similar to~\cite{Neis,Leroy} .

\bibliographystyle{unsrt}
\bibliography{report}

\end{document}